\documentclass[amsmath,amssymb,aps,twocolumn]{revtex4}

\usepackage{amsmath}
\usepackage{amssymb}
\usepackage{amsthm}
\usepackage{psfrag}
\usepackage{graphicx}

\newcommand{\be}{\begin{equation}}
\newcommand{\ee}{\end{equation}}
\newcommand{\beqa}{\begin{eqnarray}}
\newcommand{\eeqa}{\end{eqnarray}}

\def\d{\partial}
\newcommand{\bseq}{\begin{subequations}}
\newcommand{\eseq}{\end{subequations}}

\newcommand{\di}{\mathrm d}

\begin{document}
\begin{flushright}
CERN-PH-TH/2009-173
\end{flushright}
\vskip -0.9cm
\title{A healthy extension of Ho\v rava gravity}
\author{D. Blas,\!$^a$ O. Pujol\`as,\!$^b$
  S. Sibiryakov,\!$^{a,c}$}
  \affiliation{{$^a$}
 \it FSB/ITP/LPPC,
 \'Ecole Polytechnique F\'ed\'erale de Lausanne,
 CH-1015, Lausanne, Switzerland}
 \affiliation{{$^b$}\it CERN, Theory Division, CH-1211 Geneva 23,
 Switzerland}
 \affiliation{{$^c$} \it Institute for Nuclear Research of the
Russian Academy of Sciences, \\ 
      \normalsize \it  60th October Anniversary Prospect, 7a, 117312
      Moscow, Russia}

\begin{abstract}
We propose a natural extension of Ho\v rava's model for quantum gravity, which is
free from the notorious pathologies of the original proposal.
The new model endows the scalar graviton mode with a regular quadratic action and remains power-counting renormalizable. At low energies, it reduces to a Lorentz-violating scalar-tensor gravity theory. The deviations with respect to general relativity can be made weak by an appropriate choice of parameters.
\end{abstract}

\maketitle

{\em Introduction:}
%
Recently, Ho\v rava has proposed an interesting 
new approach  to
quantum gravity \cite{Horava:2009uw}. The key 
idea of the proposal is to abandon the fundamental role of the local 
Lorentz
invariance
and to assume instead that  this appears only at low energies
as an approximate symmetry. The breaking of Lorentz
invariance is achieved by equipping the space-time with an additional
geometric structure: a preferred foliation by 3-dimensional space-like
surfaces, which defines the splitting of the coordinates into space and
  time.
This allows to complete the action of 
Einstein's general relativity (GR) with higher spatial
derivatives of the metric which improve the UV behavior of
the graviton propagator and make the theory renormalizable by
power-counting. At the same time the action remains second order in
time derivatives thus avoiding the problems with ghosts which appear
in the covariant formulations of higher-derivative gravity
\cite{Stelle:1977ry}. 

The concrete realization of this idea developed in
\cite{Horava:2009uw} is formulated as follows.
One considers the ADM
decomposition of the space-time metric in the preferred foliation,
\[
\di s^2=(N^2-N_i N^i) \di t^2-2N_i  \di x^i \di t-\gamma_{ij}\di x^i \di x^j\;,
\]
and writes a generic action of the form \footnote{The 3-dimensional indexes
$i,j,\ldots$ are raised and lowered using
$\gamma_{ij}$, and covariant derivatives are associated to
$\gamma_{ij}$.}
\be
\label{ADMact}
S=\frac{M_P^2}{2}\int \di^3x \di t \sqrt{\gamma}\,N\,\big(K_{ij}K^{ij}-
\lambda K^2 -{\cal V}[\gamma_{ij}]\big)\;,
\ee
where $M_P$ is the Planck mass; $K_{ij}$ is the extrinsic curvature tensor
\be
\label{extr}
 K_{ij}=\frac{1}{2N}\left(\dot\gamma_{ij}-\nabla_i N_j-\nabla_j N_i\right)\;,
\ee
with trace $K$; 
$\gamma$ is the determinant of the
spatial metric $\gamma_{ij}$, and $N$ is the lapse function; $\lambda$
is a dimensionless constant. 

The ``potential'' term ${\cal V}[\gamma_{ij}]$ in (\ref{ADMact})
depends only on the 3-dimensional metric and its spatial derivatives and is
invariant under 3-dimensional diffeomorphisms. Explicitly,
\be
\begin{split}
\label{potential}
{\cal V}=-\xi R &+M_P^{-2}(A_1\Delta R+A_2R_{ij}R^{ij}+\ldots)\\
&+M_P^{-4}(B_1\Delta^2 R+B_2R_{ij}R^{jk}R_k^i+\ldots)\;,
\end{split}
\ee
where $R_{ij}$, $R$ are the Ricci tensor and the scalar 
curvature constructed out of the
metric $\gamma_{ij}$; $\Delta\equiv\gamma^{ij}\nabla_i\nabla_j$, and
$\xi$, $A_n$, $B_n$ are constants. 
The
ellipses represent other possible  operators 
of dimension 4 and 6
which can be constructed out of the metric $\gamma_{ij}$ and
are invariant under 3-dimensional diffeormorphisms.
As discussed
in \cite{Horava:2009uw}, restricting to the operators of dimensions up to 6 is
sufficient to make the theory 
na\"ively renormalizable by power-counting. 
In what follows we set $\xi=1$, which can always be achieved by a
suitable 
rescaling of time.

The action (\ref{ADMact}) reduces to that of GR 
if $\lambda=1$ and the terms of dimension 4 and 6 in $\cal V$ vanish. 
For other choices of parameters the model 
explicitly
breaks the general covariance down to the subgroup consisting of 
(time-dependent) 3-dimensional diffeomorphisms and 
time reparameterizations,
\be
\label{symm}
{\bf x}\mapsto\tilde{\bf x}(t, {\bf x})~,~~~t\mapsto\tilde t(t)\;,
\ee 
with the standard transformation rules for the metric components.
This invariance fixes the kinetic part of the action (\ref{ADMact}) to
be a function of $K_{ij}$. 
The symmetry (\ref{symm}) allows (but does not  require) to 
restrict $N$ to depend only on time,
$N=N(t)$. In this way one obtains the so called ``projectable''
version of the theory. Alternatively, in the
``non-projectable'' version the lapse is a generic function of space-time.
 
The introduction of terms with higher spatial derivatives in the action
leads to different 
scaling dimensions of space and time in the UV. 
Assigning dimensions
$-1$ to the space coordinates and $-3$ to time 
results in a theory
which is na\" ively power-counting renormalizable \cite{Horava:2009uw}. 
Namely, one considers the scaling transformations
\bseq
\label{scaling}
\begin{gather}
\label{scaling1}
{\bf x}\mapsto b^{-1}{\bf x}~,~~~t\mapsto b^{-3}t\;,\\
\label{scaling2}
N\mapsto N~,~~~N_i\mapsto b^{2}N_i~,~~~\gamma_{ij}\mapsto\gamma_{ij}\;.
\end{gather}
\eseq
Under this scaling, the kinetic part of the action (\ref{ADMact})
and the operators of dimension 6 in ${\cal V}$ are left unchanged (they are
marginal)
\footnote{This is true classically.
At the quantum level one expects the coefficients in front of
 marginal operators to acquire logarithmic running  under the renormalization
group flow.}.    
The rest of operators in ${\cal V}$ are relevant
deformations. 
According to the standard arguments, the action constructed from such
operators is perturbatively renormalizable. At low energies the potential is dominated by the operator
of the lowest dimension, namely, 
the spatial curvature $R$. This leads
to the recovery in the infrared 
of the relativistic scaling dimension $-1$ for both space
and time. 

At low energies, the resulting action differs from that of GR only
by the presence of the parameter $\lambda$. This would suggest that the
theory might have GR as its low-energy limit, 
provided that $\lambda$ flows to
its GR value $\lambda=1$ in the infrared. 
However, this argument has an important caveat. As already
pointed out in \cite{Horava:2009uw}, the explicit breaking of general
covariance by the preferred foliation of space-time introduces 
a new scalar degree of freedom in addition to the usual  
helicity-2 polarizations of the graviton. The subsequent study of
the properties of this extra mode has revealed that 
it persists down to low energies and exhibits a pathological behavior
which invalidates the consistency of the theory based on the action 
(\ref{ADMact}). The pathologies
include strong coupling at energies above a very low energy scale, and
fast instabilities, and appear both in the non-projectable 
\cite{Charmousis:2009tc,Li:2009bg,Blas:2009yd} 
and projectable \cite{Blas:2009yd,prep} cases. These problems can be traced back 
to
the anomalous structure of the quadratic action for the new scalar mode
around smooth backgrounds, such as Minkowski space-time. 

It was suggested in  \cite{Blas:2009yd} that the extra mode may acquire a
regular quadratic Lagrangian if the action (\ref{ADMact}) is
supplemented by certain type of new terms.
The purpose of this Letter is to show this explicitly. \\

{\em Improved behavior of the extra mode:}
Let us focus on the non-projectable version of the Ho\v rava model
and consider the following 
3-vector
\be
\label{ai}
a_i\equiv \frac{\d_iN}{N}\;.
\ee
From the geometrical point of view, $a_i$ describes the proper
acceleration of the vector field of unit normals to the foliation
surfaces \cite{Blas:2009yd}. This vector is 
manifestly covariant 
under the transformations (\ref{symm}).
This means that the potential appearing in (\ref{ADMact}) can be
extended to include terms depending on 
$a_i$ \footnote{In fact, the addition
of these terms is compulsory as nothing prevents them from being
generated by perturbative quantum corrections.}. 
Clearly, $a_i$ has dimension 1 with
respect to the scaling (\ref{scaling}). 
The requirement of
power-counting renormalizability  allows to add to the potential
(\ref{potential}) operators of dimensions up to 6, i.e. a new piece
of the form,
\begin{align}
&\delta\,{\cal V}[\gamma_{ij},a_i]=-\alpha\, a_ia^i
\label{potadd}\\
&+M_P^{-2}(C_1a_i\Delta a^i+
C_2(a_ia^i)^2+C_3 a_ia_jR^{ij}+\ldots)\notag\\
&+M_P^{-4}(D_1a_i\Delta^2 a^i+
D_2(a_ia^i)^3+D_3 a_ia^ia_ja_kR^{jk}+\ldots)\;.\notag
\end{align}
Note that the operators with odd dimensions are forbidden by spatial
parity. Similarly, the terms in the action with one time derivative of
$a_i$ are excluded by the time-reversal invariance. Finally, the terms
with two and more time derivatives acting on $a_i$ have dimension
larger than 6 and hence are not allowed by
power counting. 

We shall now demonstrate that the addition of terms of the type (\ref{potadd}) 
to the action (\ref{ADMact}) endows the extra scalar mode
 with a healthy quadratic action
at all energy scales. Let us consider the
flat metric background. Upon integrating by
parts and using Bianchi identities there are  10 inequivalent
terms in the potential that contribute to the quadratic
Lagrangian: 
\bseq
\label{oper}
\begin{align}
\label{oper1}
&(\mathrm{dim}~2)~~~~ R,~a_ia^i\;,\\
\label{oper2}
&(\mathrm{dim}~4)~~~~ R_{ij}R^{ij},~R^2,~
R\nabla_i a^i,~a_i\Delta a^i\;,\\
\label{oper3}
&(\mathrm{dim}~6)~~~~ 
(\nabla_iR_{jk})^2,~(\nabla_i R)^2,~\Delta R\, \nabla_i a^i,~
a_i \Delta^2 a^i\;.
\end{align}
\eseq  
The presence of a new operator of dimension 2 with
respect to the model of \cite{Horava:2009uw} will play a key role
in the consistency of the theory at low energies.
Introducing  the scalar 
perturbations of the metric,
\begin{align}
&N=1+\phi~,~~~~N_i=\frac{\d_i}{\sqrt{\Delta}}B\;,\\
&\gamma_{ij}=\delta_{ij}
-2\left(\delta_{ij}-\frac{\d_i\d_j}{\Delta}\right)\psi
-2\frac{\d_i\d_j}{\Delta}E\;,
\end{align} 
into the action, we obtain the
following quadratic Lagrangian for the scalar sector,
\begin{align}
{\cal L}^{(2)}&=\frac{M_P^2}{2}\bigg\{-2\dot\psi^2
-2\psi\Delta\psi+4\phi\Delta\psi
+4\psi\sqrt\Delta\dot B\notag\\
&+4\psi\ddot E
-(\lambda-1)\left(\sqrt\Delta B+
\dot E+2\dot\psi\right)^2
+\alpha(\d_i\phi)^2\notag\\
&-\frac{f_1}{M_P^2}(\Delta\psi)^2-\frac{2f_2}{M_P^2}\Delta\phi\Delta\psi
-\frac{f_3}{M_P^2}(\Delta\phi)^2\notag\\
&-\frac{g_1}{M_P^4}\psi\Delta^3\psi
-\frac{2g_2}{M_P^4}\phi\Delta^3\psi
-\frac{g_3}{M_P^4}\phi\Delta^3\phi\bigg\}\;,\label{L2imp}
\end{align}
where the constants $\alpha$, $f_n$, $g_n$ are related to the
coefficients in front of the operators (\ref{oper}) in the potential.
In particular, $\alpha$ is the coefficient in front of the 
operator $a_ia^i$.
Fixing 
the gauge $B=0$
and integrating out 
the non-dynamical fields 
$E$  and $\phi$,  we obtain
\be
\label{Lscalar}
\begin{split}
{\cal L}^{(2)}=\frac{M_P^2}{2}\bigg\{
\frac{2(3\lambda-1)}{\lambda-1}\dot\psi^2
+\psi\,\frac{P[M_P^{-2}\Delta]}{Q[M_P^{-2}\Delta]}\Delta \psi\bigg\}\;,
\end{split}
\ee
where the polynomials $P$, $Q$ have the form,
\begin{align}
P[x]=&(g^2_2-g_1g_3)x^4-(g_1f_3+g_3f_1-2g_2f_2)x^3\notag\\
&+(f_2^2-4g_2-f_1f_3-2g_3-g_1\alpha)x^2\notag\\
&-(2f_3+f_1\alpha+4f_2)x+(4-2\alpha)\;,
\label{poly1}\\
\label{poly2}
Q[x]=&g_3x^2+f_3x+\alpha\;.
\end{align}
The Lagrangian (\ref{Lscalar}) describes a healthy excitation provided
that two conditions
are satisfied. First, to avoid ghosts, the time-derivative term
must be positive definite. This puts a constraint on the parameter
$\lambda$,
\be
\label{lambdacons}
\frac{3\lambda-1}{\lambda-1}>0\;.
\ee
This condition can be easily fullfilled, e.g. by choosing $\lambda>1$. 
Second, from  the dispersion
relation of the propagating mode $\psi$,
\be
\label{disp}
\omega^2=\frac{\lambda-1}{2(3\lambda-1)}\;
\frac{P[-p^2/M_P^2]}{Q[-p^2/M_P^2]}\;p^2\;,
\ee
one reads off the condition to avoid exponential 
instabilities 
(assuming that (\ref{lambdacons}) holds),
\be
\label{condPQ}
P[x]/Q[x]>0~~~~~~ \mathrm{ at}~~ x<0\;.
\ee
This condition puts certain restrictions on the constants $\alpha$,
$f_n$, $g_n$. In particular, we obtain that $\alpha$ must belong to
the interval
\be
\label{alphagood}
0<\alpha<2\;.
\ee
The precise form of the constraints on the other parameters
coming from (\ref{condPQ})
 is quite cumbersome and we prefer to omit it in this Letter. Nevertheless, the
reader can easily convince himself that there is a non-empty region
of the parameter space where (\ref{condPQ}) is satisfied.

In deriving (\ref{Lscalar}) we have used in an essential
way the dependence of the potential of the model on $a_i$. Indeed, in
the absence of such dependence, as happens in the non-projectable
version of the
original Ho\v rava's proposal, the constants $\alpha$, $f_2$, $f_3$,
$g_2$, $g_3$ become zero and the polynomial $Q[x]$ vanishes  identically. This
means that the  Lagrangian 
(\ref{Lscalar}) is singular in this limit.   

We can also compare the situation in our model with the projectable case of
Ho\v rava gravity. The latter is obtained from our expressions by
taking the limit  
$\alpha\to\infty$ which forces $\phi$ (the perturbation of $N$) to be constant in space. From
the dispersion relation (\ref{disp}) one reads that in this case the
scalar mode has an imaginary sound speed at low energies, 
cf. \cite{Sotiriou:2009bx},
$$
c_{proj}^2= - \frac{\lambda-1}{3\lambda-1} <0\;.
$$ 
This leads to an exponential instability, that can be tamed only 
by the higher order terms in the dispersion relation. Thus, the 
characteristic rate of the instability is of order
$|c_{proj}| M_P$. In principle, this rate can be suppressed by
choosing $(\lambda-1)$ to be extremely small. However, in this case the
strong coupling scale of the theory becomes unacceptably low
\cite{Blas:2009yd,prep}. 

Let us return to our model.
It is important to stress that the healthy behavior of the scalar mode
can be achieved simultaneously with the stability in the sector of the
helicity-2 perturbations. Indeed, the dispersion relation for the
latter depends only on the coefficients in front of the 
operators in the first
column of the list (\ref{oper}). After fixing these coefficients to
ensure stability of the helicity-2 modes, we
still have the freedom to choose the coefficients of the remaining 
operators in the list. This amounts to the possibility of freely choosing
the constants $\alpha$, $f_n$, $g_n$ in the scalar Lagrangian
(\ref{L2imp}) to satisfy (\ref{condPQ}).  

The existence of a healthy quadratic action for the perturbations around
Minkowski space-time guarantees the absence of short-scale
instabilities for any smooth background. Indeed, at short scales a
smooth metric can be approximated by the flat one, implying that the
short-wavelength perturbations around this metric behave in the same
way as in Minkowski. Additionally, a regular quadratic Lagrangian
allows to develop the standard perturbation theory to account for the
interactions of the modes. Together with the power-counting
renormalizability of the model, this strongly suggests that, with 
appropriate choice of parameters, the theory is
free of strong coupling at all energies.
Still, an explicit analysis 
of the perturbation theory series
is needed to check this conjecture. \\

{\em Newton's law and low energy cosmology:}
%
At low energies the dispersion relation (\ref{disp}) for the scalar
mode becomes linear:
\be
\label{dispimp}
\omega^2=\frac{\lambda-1}{3\lambda-1}\left(\frac{2}{\alpha}-1\right)p^2\;.
\ee
Depending on the values of $\lambda$ and $\alpha$, the
propagation velocity of the scalar may differ from 1 (the velocity of
the helicity-2 modes), which
means that  Lorentz
invariance is generically broken at low energies. The presence of a
gap-less scalar gravitational mode potentially implies an interesting
low-energy phenomenology of the model to be confronted with the
existing tests of
GR \cite{Will:2005va}. Leaving a comprehensive study of this issue
for future research, we analyze in this section
two basic phenomenological aspects
of the model. 

First, we consider the large distance behaviour of the
gravitational field of a static point-like
source of mass $m$.
Note that only the scalar part of the metric is excited in this case. 
The corresponding low-energy Lagrangian
is obtained by combining the first two
lines of Eq.~(\ref{L2imp}) with the source term. 
The static part of the
Lagrangian is given by 
\be
\label{Lint}
{\cal L}=\frac{M_P^2}{2}\Big(-2\psi\Delta\psi
+4\phi\Delta\psi+\alpha(\d_i\phi)^2\Big)-m\phi\delta^3({\bf x})\;.
\ee 
The solution of the equations of motion following from this Lagrangian
reads,
\be
\label{Newton}
\phi=\psi=-\frac{m}{8\pi M_P^2(1-\alpha/2)|{\bf x}|}\;.
\ee
Remarkably, the gravitational field has the same form as in GR
with the effective Newton constant 
\be
\label{GN}
G_N=\frac{1}{8\pi M_P^2(1-\alpha/2)}\;.
\ee
In particular, the relation (\ref{Newton})
implies that, in contrast to the case of Lorentz-invariant
scalar-tensor theories of gravity \cite{Will:2005va}, the deflection of light by the
gravitational field in our model is the same as in GR. 

The second phenomenological aspect that we discuss
is low-energy cosmology.
Notice that for the spatially homogeneous
metric ansatz the proper acceleration $a_i$ vanishes. As a
consequence, the evolution of the Universe is insensitive to the terms
with $a_i$ in the action and differs at large distances from the
  case of GR only due to the presence of the parameter
  $\lambda$. Substituting the FRW metric into the action and varying
  with respect to the lapse one obtains the standard Friedmann
  equation, 
\be
\label{Friedman}
H^2=\frac{8\pi}{3}\,G_{cosm}\,\rho\;,
\ee 
where $H$ is the Hubble parameter, $\rho$ is the total density of the
Universe. The effective gravitational constant is
\be
\label{Gcosm}
G_{cosm}=\frac{2}{8\pi M_P^2(3\lambda-1)}\;.
\ee
Note that $G_{cosm}\neq G_N$. A similar discrepancy between the
gravitational constants appearing in the Newton's law and in the
Friedmann equation also arises in certain low-energy
theories constructed to describe the Lorentz-violating effects in
gravity. These models include the Einstein--aether theory (see
\cite{Jacobson:2008aj} for a recent review) and the gauged ghost
condensate  \cite{Cheng:2006us}. The observational bound on this
discrepancy comes from the measurement of the primordial abundance of
He$^4$ and reads \cite{Carroll:2004ai,Jacobson:2008aj} 
$|G_{cosm}/G_N-1|\lesssim 0.13$.
In our model this implies rather mild constraints on the 
parameters $\alpha$ and $\lambda$.\\

{\em Discussion:}
%
In this Letter we have described a  
natural extension 
of the non-projectable version of Ho\v rava's proposal
for quantum gravity, which is free from the pathologies present in the
original formulation of that proposal. The extension  is obtained
  by including in the action all terms allowed by the symmetries and the
requirement that the model is power-counting renormalizable. 
It remains to be seen if the proposed model  provides a valid
theory of quantum gravity. 

At low energies the model does
not reduce to GR but to a Lorentz-violating scalar--tensor theory. 
This potentially implies a rich low-energy phenomenology to
be confronted with existing tests of GR.
Remarkably, the effects of the scalar mode at large
distances can be made weak by an appropriate choice of the parameters
without spoiling the good features of the model. 
It is clear that a detailed study of the 
phenomenological aspects of the theory will provide further
constraints on its parameters.

A common problem of any theory
with high-energy breaking of Lorentz symmetry is
the mechanism to 
recover the Lorentz
invariance in the infrared.
This issue arises because the
Lorentz violation in the UV generically translates at low energies into
the difference of the limiting propagation velocities for 
different particle
species \cite{Collins:2004bp}. On the other hand, such differences are
tightly constrained experimentally \cite{Mattingly:2005re}. 
This seems to require a very precise fine-tuning of parameters 
to reconcile the theory with
experiment. A more elegant solution
would be to invoke some (super)symmetry
which relates all the matter species in the UV and is broken at a
scale much lower than the characteristic scale of Lorentz violation
(which in the present setup is naturally of order the Planck
mass). In this case, the Lorentz violating effects would be suppressed at low
energies by the ratio of the two scales.  Clearly, 
this remains an open issue.  

At the purely theoretical level,
the violation of Lorentz invariance down to the infrared
leads to  apparent paradoxes even in the
absence of any matter fields.
We have seen that
the propagation velocity of the scalar gravitational waves is in
general different from that of the helicity-2 modes. 
This opens up the possibility to realize a
perpetuum mobile of the second kind, and hence to violate unitarity, 
in gedanken processes involving black holes 
\cite{Dubovsky:2006vk,Eling:2007qd}. It will be interesting to see if
and how this puzzle is resolved in the model proposed in this Letter.

\paragraph*{Acknowledgments}

We thank Robert
Brandenberger, Shinji Mukohyama and Igor Tkachev for stimulating discussions.  
This work was supported in part by the Swiss Science Foundation
(D.B.), the Tomalla Foundation (S.S.), RFBR grant 08-02-00768-a (S.S.)
and the Grant of the President of Russian Federation
NS-1616.2008.2 (S.S.).


\end{document}